\documentclass{acm_proc_article-sp}

\usepackage{xcolor}
\usepackage[final]{listings}
\usepackage{amsmath, amssymb}
\usepackage{mathtools}
\usepackage{hyperref}
\usepackage{tikz}
\usepackage{pgfplots}
\usepackage{xspace}

\usepackage{aicescover}

\colorlet{graybg}{gray!10}
\colorlet{plot1}{red}
\colorlet{plot2}{green!75!black}
\colorlet{plot3}{blue}
\colorlet{plot4}{yellow!75!black}
\colorlet{plot5}{violet}
\colorlet{plot6}{cyan}

\lstdefinelanguage{algs}{
    keywords=[1]{for,each,in,end,set,if,not},
    keywordstyle=[1]{\bf},
    keywords=[2]{load_start,load_wait,store_start,store_wait},
    keywordstyle=[2]{\bf\green},
    keywords=[3]{copy},
    keywordstyle=[3]{\bf\blue}
}
\usetikzlibrary{
    external,
    scopes,
    patterns,
    calc,
    plotmarks,
    fpu
}
\pgfplotsset{
    every axis/.append style={
        axis background/.style={
            fill=graybg
        },
        legend cell align=left,
        xmin=0,
        ymin=0,
        xlabel near ticks,
        ylabel near ticks,
        enlarge x limits={
            value=0.05,
            auto
        },
        enlarge y limits={
            value=0.05,
            auto
        },
        scaled ticks=false,
        width=.5\textwidth,
        height=.3\textwidth,
        ymajorgrids=true,
    },
    every axis plot/.append style={
        thick
    },
}
\pgfkeys{
    /tikz/external/mode=list and make
}
\tikzsetexternalprefix{figures/externalized/}
\tikzset{external/export=false}

\let\oldtheequation\theequation
\makeatletter
\def\tagform@#1{\maketag@@@{\ignorespaces#1\unskip\@@italiccorr}}
\renewcommand{\theequation}{(\oldtheequation)}
\makeatother

\hypersetup{
    colorlinks=true,       % false: boxed links; true: colored links
    linkcolor=blue!50!black,          % color of internal links
    citecolor=green!50!black,        % color of links to bibliography
    filecolor={.5 0 0},      % color of file links
    urlcolor=red!50!black           % color of external links
}

\newcommand{\red}[1]{\textcolor{red!75!black}{#1}}
\newcommand{\green}[1]{\textcolor{green!75!black}{#1}}
\newcommand{\blue}[1]{\textcolor{blue}{#1}}
\newcommand{\splitto}[2]{\biggl(\!\begin{array}{c} #1 \\\hline #2 \end{array}\!\biggr)}
\newcommand{\splitot}[2]{\bigl(#1 \big| #2 \bigr)}
\newcommand{\splittt}[4]{\biggl(\!\begin{array}{c|c} #1 & #2 \\\hline #3 & #4 \end{array}\!\biggr)}

\newcommand{\smpooc}{{\sc SMP-OOC}\xspace}
\newcommand{\smpic}{{\sc SMP-IC}\xspace}
\newcommand{\distooc}{{\sc Elem-OOC}\xspace}

\newcommand{\pdj}[1]{\textcolor{blue}{\bf Pdj: #1}}
\newcommand{\ep}[1]{\textcolor{green}{\bf EP: #1}}

\title{Algorithms for Large-scale\\ Whole Genome Association Analysis}

\numberofauthors{4}
\author{
    \alignauthor
    Elmar Peise\\
    \affaddr{RWTH Aachen University}\\
    \affaddr{Aachen, Germany}\\
    \email{\small \tt peise@aices.rwth-aachen.de}
    \alignauthor
    Diego Fabregat\\
    \affaddr{RWTH Aachen University}\\
    \affaddr{Aachen, Germany}\\
    \email{\small \tt fabregat@aices.rwth-aachen.de}
    \alignauthor
    Yurii Aulchenko\\
    \affaddr{Institute of Cytology and Genetics}\\ 
    \affaddr{Novosibirsk, Russia}
    \and
    \alignauthor
    Paolo Bientinesi\\
    \affaddr{RWTH Aachen University}\\
    \affaddr{Aachen, Germany}\\
    \email{\small \tt pauldj@aices.rwth-aachen.de}
}

\aicescoverauthor{
    Elmar Peise
    \and
    Diego Fabregat
    \and
    Yurii Aulchenko
    \and
    Paolo Bientinesi
}

\begin{document}
\aicescoverpage
\maketitle
\begin{abstract}
    In order to associate complex traits with genetic polymorphisms,
    genome-wide association studies process huge datasets involving tens of
    thousands of individuals genotyped for millions of polymorphisms. 
    When handling these datasets, which exceed 
    the main memory of contemporary computers,
    one faces two distinct challenges: 
    1) Millions of polymorphisms come at the cost of
    hundreds of Gigabytes of genotype data,
    which can only be kept in secondary storage; 
    2) the relatedness of the test population is
    represented by a covariance matrix, which, for large populations, can only
    fit in the combined main memory of a distributed architecture.
    In this paper, we present solutions for both challenges:
    The genotype data is streamed from and to secondary storage 
    using a double buffering technique,
    while the covariance matrix is kept across the main memory of a distributed
    memory system.  
    We show that these methods sustain high-performance 
    and allow the analysis of enormous datasets.
\end{abstract}


\section{Introduction}
Whole Genome Association Studies, 
also known as Genome-Wide Association (GWA) studies, 
became the tool of choice for the
identification of loci associated with complex traits.
The association between a trait of interest and genetic polymorphisms (usually
single nucleotide polymorphisms, SNPs) is studied using thousands of people
typed for hundreds of thousands of polymorphisms.
Thanks to these studies,
hundreds of loci for dozens of complex human diseases and quantitative
traits have been discovered~\cite{pmid19474294}. 
In GWA analysis, one of the most used methods to account for the genetic substructure due
to relatedness and population stratification is the variance component
approach based on mixed models~\cite{pmid3435047,pmid16380716}. 
While effective, mixed-models based methods are computationally demanding
both in terms of data management and computation.
The objective of this research is to make large-scale GWA analyses more affordable.

Computationally, a GWA analysis 
based on approximations to the mixed-model applied to a set of 
$n$ individuals and $m$ genetic markers (SNPs)
boils down to the solution of $m$ generalized least-squares (GLS) problems 
\begin{equation}
\label{eq:GWAS}
b_i \coloneqq \bigl( X_i^T M^{-1} X_i \bigr)^{-1} X_i^T M^{-1} y, 
\ \ \text{with} \ \  i=1, \dots, m,
\end{equation}
where the $X_i \in \mathbb{R}^{n \times p}$ is the design matrix,
$M \in \mathbb{R}^{n \times n}$ is the covariance matrix, 
the vector $y \in \mathbb{R}^{n}$ contains the phenotypes, and
the vector $b_{i} \in \mathbb{R}^{p}$ 
expresses the relation between a variation in the SNP ($X_i$)
and a variation in the trait ($y$).
Additionally, $M$ is symmetric positive definite (SPD), 
$2 \le p \le 20$, $n$ ranges approximately between $10^3$ and $10^5$,
and $m$ ranges between $10^5$ and $10^8$.
Finally, $X_i$ is full rank and can be viewed as composed of two parts,
$X_i = \splitot{X_L}{X_{Ri}}$, with $X_L \in \mathbb{R}^{n \times (p-1)}$
and $X_{Ri} \in \mathbb{R}^{n \times 1}$, where $X_L$ is constant across all
$m$ genetic markers.

The first reported GWA study dates back to 2005: 
$146$ individuals were genotyped, and about $103{,}000$ SNPs were analyzed~\cite{firstgwas}.
Since then,
as the catalog of publishd GWA analyses shows~\cite{gwascatalog, gwastrend},
the number of published studies has increased steadily, 
up to $2{,}404$ in 2011 and $3{,}307$ in 2012.
A similar growth can be observed 
both in the population size and in the number of SNPs:
Across all the GWAS published in 2012, 
on average, the studies used $15{,}471$ individuals, 
with a maximum of $133{,}154$,
and $1{,}252{,}222$ genetic markers, 
with a maximum of $7{,}422{,}970$.
From the perspective of~\autoref{eq:GWAS}, 
these trends present concrete challenges,  
especially in terms of memory requirements.
As both  $M \in \mathbb R^{n \times n}$ and the $X_i$'s compete for the main
memory, two scenarios arise:
1) If $n$ is small enough, $M$ fits in memory, and the $X_i$'s have to be
streamed from disk; 2) if $M$ does not fit in memory, then both data
and computation have to be distributed over multiple compute nodes.
In this paper we present efficient algorithms for both scenarios. 

Several notable implementations for GWA studies already exist:
{\sc GenABEL} is a widely spread library for genome studies~\cite{GenABEL};
{\sc FaST-LMM} is a program specifically designed for large datasets~\cite{FastLMM};
recently, a new high-performance implementation, to which we refer as \smpooc, 
was introduced in~\cite{diego}.
None of these algortihms are meant for distributed memory systems;
hence, for all of them the population size $n$ is 
limited by the memory of a single node.

The rest of this paper is structured as follows. 
A mathematical algorithm used to solve~\autoref{eq:GWAS} 
is discussed in~\autoref{sec:alg}.
Then, a technique to make the algorithm feasible for 
an arbitrary numbers of SNPs out-of-core is presented in~\autoref{sec:diego}.
Finally, 
in~\autoref{sec:dist},
the algorithm is further extended to deal with large population sizes 
by means of distributed memory architectures.

%%% Local Variables: 
%%% mode: latex
%%% TeX-master: "Main"
%%% End: 

\section{The Mathematical Algorithm}
\label{sec:alg}
The standard route to solving one of the GLS's in \autoref{eq:GWAS} is to
reduce it to an ordinary least squares problem (OLS),
$$
    b_i = \bigl(\overline X_i^T \overline X_i\bigr)^{-1} \overline y,
$$
through the operations
\begin{lstlisting}
$L L^T \coloneqq M$ !%
!               !(Cholesky factorization)!
$\overline X_i \coloneqq L^{-1} X_i$ !%
!               !(triangular solve)!
$\overline y \coloneqq L^{-1} y$ !%
!               !(triangular solve)!
\end{lstlisting}

The resulting OLS can then be solved by two alternative approaches,
respectively based on the QR decomposition of  $\overline X_i$,
and the Cholesky decomposition of $\overline X_i^T \overline X_i$.  
In general, the QR-based method is numerically more stable; 
however, in this specific application, 
since $\overline X_i^T \overline X_i \in \mathbb R^{p \times p}$ is very small
and $X$ is typically well conditioned, both approaches are
equally accurate. 
In terms of performance, the solution via Cholesky decomposition (detailed
below) is slightly more efficient. 
\begin{lstlisting}[firstnumber=last]
$S_i \coloneqq \overline X_i^T \overline X_i$ !%
!               !(symmetric matrix product)!
$\overline b_i \coloneqq \overline X_i^T \overline y$ !%
!               !(matrix times vector)!
$b_i \coloneqq S_i^{-1} \overline b_i$ !%
!               !(linear system via Cholesky)!
\end{lstlisting}
In this paper, we only consider this approach.

\subsection{Multiple SNPs}
When the six steps for the solution of one OLS are 
applied to the specific case of \autoref{eq:GWAS}, it is possible to
take advantage of the structure of $X_i$ and
avoid redundant computation.

Plugging $X_i = \splitot{X_L}{X_{Ri}}$ into
$\overline X_i \coloneq L^{-1} X_i$ (line {\tt 2}),
we obtain 
$$
\splitot{\overline X_L}{\overline X_{Ri}} 
\coloneqq \splitot{L^{-1} X_L}{L^{-1} X_{Ri}},
$$
that is,
$\overline X_L \coloneqq L^{-1} X_L$, and 
$\overline X_{Ri} \coloneqq L^{-1} X_{Ri}$. 
These assignments indicate that the quantity $\overline X_L$ can be
computed once and reused across all the SNPs.

Similarly, for $S_i \coloneqq \overline X_i^T \overline X_i$ (line {\tt 4}),
we have\footnote{The subscript letters $_L$, $_R$, $_T$, and $_B$ 
  stand for Left, Right, Top, and Bottom, respectively.}
$$
\splittt{S_{TL}}{\ast}{S_{BLi}}{S_{BRi}} 
\coloneqq 
\splittt{\overline X_L^T \overline X_L}{\ast}
{\overline X_{Ri}^T \overline X_L}{\overline X_{Ri}^T \overline X_{Ri}},
$$
from which
\begin{eqnarray}
    S_{TL} &\coloneqq &\overline X_L^T \overline X_L \in \mathbb R^{(p - 1) \times (p - 1)}, \nonumber\\
    S_{BLi} &\coloneqq &\overline X_{Ri}^T \overline X_L^T \in \mathbb R^{1 \times (p - 1)}, \text{ and } \nonumber \\
    S_{BRi} &\coloneqq &\overline X_{Ri}^T \overline X_{Ri}^T \in \mathbb R, \nonumber
\end{eqnarray}
indicating that $S_{TL}$, the top left portion of $S_i$, is independent of $i$
and needs to be computed only once.%
\footnote{Since $S_i$ 
  is symmetric, its top-right and bottom-left quadrants are 
  the transpose of each other; we mark the top-right quadrant with a $\ast$,
  indicating that it is never accessed nor computed.} 
Finally, the same idea applies also to $\overline b_i$ (line {\tt 5}),
yielding the assignments
$\overline b_T \coloneqq \overline X_L^T y$ 
and
$\overline b_{Bi} \coloneqq \overline X_{Ri} y$.

\begin{lstlisting}[
    float=t,
    label=alg:base,
    caption={
        Optimized algorithm for the solution of~\autoref{eq:GWAS}.
    }
]
  $L L^T \coloneqq M$
  $\overline X_L \coloneqq L^{-1} X_L$, !%
  !           $\overline y \coloneqq L^{-1} y$
  $S_{TL} \coloneqq \overline X_L^T \overline X_L$, !%
  !           $\overline b_T \coloneqq \overline X_L^T y$
  for $i$ in $\{1, \ldots, m\}$
      $\overline X_{Ri} \coloneqq L^{-1} X_{Ri}$
      $S_{BLi} \coloneqq \overline X_{Ri}^T \overline X_L$
      $S_{BRi} \coloneqq \overline X_{Ri}^T \overline X_{Ri}$
      $\overline b_{Bi} \coloneqq \overline X_{Ri}^T \overline y$
      set $S_i \coloneqq \splittt{S_{TL}}{\ast}{S_{BLi}}{S_{BRi}}$, $\overline b_i \coloneqq \splitto{\overline b_T}{\overline b_{Bi}}$
      $b_i \coloneqq S_i^{-1} \overline b_i$
  end
\end{lstlisting}

%%% Local Variables: 
%%% mode: latex
%%% TeX-master: "../Main"
%%% End: 

The computation for the whole \autoref{eq:GWAS} is given in
\autoref{alg:base}. There, 
all the operations independent of $i$ are moved outside the loop, 
thus lowering the overall complexity from 
$O(n^3 + m n^2 p)$ down to $O(n^3 + m n^2)$.\footnote{
Since in most analyses  $m \gg n$, 
the complexity reduces by a factor of $p$, 
from $O(m n^2 p)$ down to $O(m n^2)$.
}
This algorithm constitutes the basis for the large-scale 
versions presented in the next two sections.

%%% Local Variables: 
%%% mode: latex
%%% TeX-master: "Main"
%%% End: 

\section{Out-of-core}
\label{sec:diego}
GWA studies often operate on and generate 
datasets that exceed the main memory capacity of current computers.  
For instance, a study with $n = 20{,}000$ individuals, $m = 10{,}000{,}000$ SNPs, 
and $p=4$, 
requires 1.49~TB to store the input data ($M$ and $X_i$'s), and
generates 305~MB of output.\footnote{In practice the size of the
  output is even larger, because in addition to $b_i$, 
  a $p\times p$ symmetric matrix is also generated. }  
To make large analyses feasible, regardless of the number of SNPs, 
Fabregat et al.~proposed an extended version of~\autoref{alg:base}, described below,
that streams $X_{Ri}$ and $b_i$ from secondary storage,
by means of asynchronous I/O operations~\cite{diego}.

\begin{lstlisting}[
    float=t,
    label=alg:ooc,
    caption={
      Out-of-core version of~\autoref{alg:base}:
      The $X_{Ri}$ and $b_i$ are streamed from and to disk in blocks.
      Asynchronous I/O operations are in \green{green}.
        %   \pdj{ should you make clear what ``current''/''next'' means?
        % maybe you need an assignment that says ``next'' becomes ``current''?}
        % \ep{I think it's pretty clear. The assignment would mess things up by
        % mixing levels of abstraction}
    }
]
$L L^T \coloneqq M$
$\overline X_L \coloneqq L^{-1} X_L$, !%
!           $\overline y \coloneqq L^{-1} y$
$S_{TL} \coloneqq \overline X_L^T \overline X_L$, !%
!           $\overline b_T \coloneqq \overline X_L^T y$
load_start !first! $X_{blk}$
for each $blk$
    load_wait !current! $X_{blk}$
    if not !last! $blk$: load_start !next! $X_{blk}$
    $\overline X_{blk} \coloneqq L^{-1} X_{blk}$
    for $i$ in  $\{1, \ldots, m_{blk}\}$
        set $\overline X_{Ri} \coloneqq \overline X_{blk}[i]$
        $S_{BLi} \coloneqq \overline X_{Ri}^T \overline X_L$, !%
        !           $S_{BRi} \coloneqq \overline X_{Ri}^T \overline X_{Ri}$
        $\overline b_{Bi} \coloneqq \overline X_{Ri}^T \overline y$
        set $S_i \coloneqq \splittt{S_{TL}}{\ast}{S_{BLi}}{S_{BRi}}$, $\overline b_i \coloneqq \splitto{\overline b_T}{\overline b_{Bi}}$
        $b_i \coloneqq S_i^{-1} \overline b_i$
        set $b_{blk}[i] \coloneqq b_i$
    end
    if not !first! $blk$: store_wait !previous! $b_{blk}$
    store_start !current! $b_{blk}$
end
store_wait !last! $b_{blk}$
\end{lstlisting}

%%% Local Variables: 
%%% mode: latex
%%% TeX-master: "../Main"
%%% End: 

In order to avoid any overhead,
the vectors $X_{Ri}$ (and $b_i$) are grouped into blocks $X_{blk}$
(and $b_{blk}$) of size $m_{blk}$, and read (written)
asynchronously using double buffering.
The idea is to logically split the main memory in two equal regions: 
One region is devoted to the block of data that is currently processed, 
while the other is used to store the output from the previous block and to 
load the input for the next one. Once the computation on the current block is
completed, the roles of the two regions are swapped.
The algorithm commences by loading the first block of SNPs $X_{blk}$ from disk
into memory; 
then, while the GLS's corresponding to this block are solved,
the next block of SNPs is loaded asynchronously in the second memory region.
(Analogously, the previous $b_{blk}$ is stored, while the current 
one is computed.)  

When dealing with large analyses, an important optimization comes from, whenever possible,
processing multiple SNPs at once: \autoref{alg:ooc}
shows how slow vector operations on $X_{Ri}$ can be combined together,
originating efficient matrix operations on 
$X_{blk} \in \mathbb R^{n \times m_{blk}}$.

\subsection{Shared memory implementation}
\label{sec:smp}
The implementation of \autoref{alg:ooc}, called~\smpooc, 
makes use of parallelism in two different ways~\cite{diego}.
The operations in lines {\tt 1} through {\tt 8} are dominated by BLAS3 
and take full advantage of a multithreaded implementation of BLAS (LAPACK).
By contrast, 
the operations within the innermost loop 
(lines {\tt 11} through {\tt 14}), 
only involve very small or thin matrices, for which BLAS and especially multithreaded BLAS are less efficient.
Therefore, thery are scheduled in parallel using {\sc OpenMP} in
combination with single-threaded {\sc BLAS} and {\sc LAPACK}.

We compiled \smpooc,
written in C, with the GNU C~compiler (version~4.4.5)
and linked to Intel's Math Kernel Library (MKL version~10.3).
All tests were executed on
a system consisting of two six-core Intel X5675 processors, running at 3.06~GHz,
equipped with 32~GB of RAM, and connected to a 1~TB hard disk.

Preliminary measurements have shown that changing $p \in \{1, \ldots, 20\}$
results in performance variation on the order of system fluctuations (below $1
\%$). We therefore consider $p = 4$, a value encountered in several GWA
studies, throughout all our experiments.

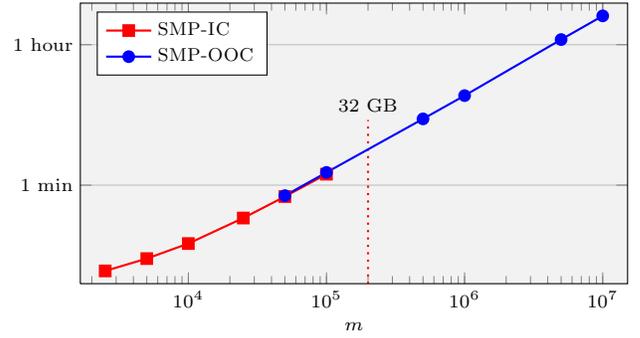
\begin{figure}[t]
    \scriptsize
    \centering
    \tikzset{external/export=true}

    \begin{tikzpicture}
        \begin{loglogaxis}[
            xlabel={$m$},
            legend pos=north west,
            ytick={60,3600},
            yticklabels={1 min, 1 hour},
        ]
            \draw[plot1, thick, dotted] (axis cs:200000, 0) -- (axis cs:200000, 400) node[anchor=south, black] {32 GB};
            \addplot[plot1, mark=square*] table[x index=0, y index=1] {figures/data/diego/hp-gwas_12th.dat};
            \addlegendentry{\smpic}
            \addplot[plot3, mark=*] table[x index=0, y index = 1] {figures/data/diego/ooc-hp-gwas_12th.dat};
            \addlegendentry{\smpooc}
        \end{loglogaxis}
    \end{tikzpicture}

    \caption[Performance of \smpic and \smpooc for increasing $m$]{
        Performance of \smpic and \smpooc for increasing $m$. The vertical line
        is the limit for the in-core version imposed by the RAM size.
        $n = 10{,}000$, $p = 4$.
    }
    \label{fig:diegoooc}
    \tikzset{external/export=false}
\end{figure}

In the first experiment, we compare the efficiency of \smpooc with \smpic, an equivalent in-core version.  
We fixed $n = 10{,}000$, $p = 4$, and we let 
$m$ vary between $10^3$ and $10^7$.
For the out-of-core version, 
the SNPs were grouped in blocks of size $m_{blk} = 5{,}000$.
\autoref{fig:diegoooc} shows that \smpooc
scales linearly in the number of SNPs, well beyond
the maximum problem size imposed by the 32~GB of RAM. 
Furthermore, the fact that the lines for the in-core and out-of-core algorithms
overlap perfectly confirms that the I/O operation to and from disk are
entirely hidden by computation.

\begin{figure}[t]
    \scriptsize
    \centering
    \tikzset{external/export=true}

    \begin{tikzpicture}
        \begin{loglogaxis}[
            xlabel={$m$},
            legend pos=south east,
            xmax=7e7,
            xtick={1e6,1e7,3.6e7},
            xticklabels={$10^6$,$10^7$,$3.6 \cdot 10^7$},
            xtick=data,
            ytick={60,3600,86400,604800},
            yticklabels={1 min, 1 hour, 1 day, 1 week},
        ]
            \addplot[plot1, mark=triangle*] table[x index=0, y index=3] {figures/data/diego/others.dat}
                node[black,anchor=west] {$\times 56.8$};
            \addlegendentry{\sc GenABEL}
            \addplot[plot2, mark=square*] table[x index=0, y index=2] {figures/data/diego/others.dat}
                node[black,anchor=west] {$\times 6.3$};
            \addlegendentry{\sc FaST-LMM}
            \addplot[plot3, mark=*] table[x index=0, y index=1] {figures/data/diego/others.dat}
                node[black,anchor=west] {$\times 1$};
            \addlegendentry{\smpooc}
        \end{loglogaxis}
    \end{tikzpicture}

    \caption[Performance of \smpooc, {\sc GenABEL}, and {\sc FaST-LMM}]{
        Performance of \smpooc compared to {\sc GenABEL} and {\sc FaST-LMM}.
        $n = 10{,}000$, $p = 4$.
    }
    \label{fig:diegoothers}
    \tikzset{external/export=false}
\end{figure}
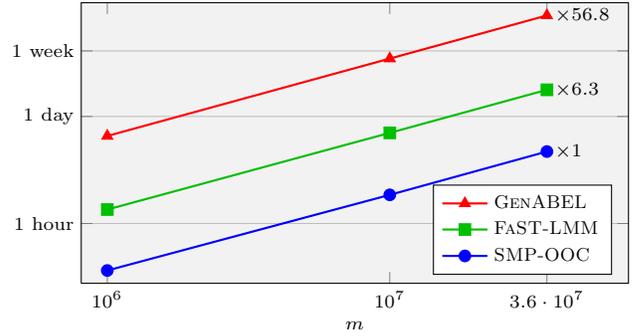

In the second experiment, \autoref{fig:diegoothers},  
we show the performance of \smpooc
with respect to that of other two solvers:
{\sc FaST-LMM}, a program designed for GWAS on large 
datasets~\cite{FastLMM} and {\sc GenABEL}, 
a widely spread library for genome studies~\cite{GenABEL}.  
Again, we fixed $n = 10{,}000$ and $p = 4$, while $m$ varies between $10^6$ and $3.6 \cdot 10^7$.
The fairly constant observed speedups of \smpooc over {\sc FaST-LMM} and {\sc
GenABEL} are, at $m = 3.6 \cdot 10^7$, $6.3$ and $56.8$, respectively.

%%% Local Variables: 
%%% mode: latex
%%% TeX-master: "Main"
%%% End: 

\section{Distributed Memory}
\label{sec:dist}
While~\smpooc scales up to an arbitrarily large amount of SNPs $m$,
the main memory is still a limiting factor for the population size
$n$: In fact, the algorithm necessitates the matrix 
$M \in \mathbb R^{n \times n}$ (or equivalently, its 
Cholesky factor $L$) to reside fully in memory.
Due to the triangular solve (\autoref{alg:ooc}, line {\tt 2}),
keeping the matrix in the secondary storage is not a viable
option. 
Our approach here consists in distributing 
$M$, $L$, and all matrices on which $L$ operates,
across multiple compute nodes, lifting any constraint on their size.

\subsection{Elemental}
\label{sec:elemental}

\begin{figure}[t]
    \scriptsize
    \centering
    \tikzset{external/export=true}

    \begin{tikzpicture}
        \node[matrix, plot1, label={[inner sep=0]above:$p_0$:}] (p0) {
            \node {$a_{11}$}; & \node {$a_{14}$}; \\
            \node {$a_{31}$}; & \node {$a_{34}$}; \\
        };
        \path (p0.south) ++(0, -.2) node[matrix, plot2, anchor=north, label={[inner sep=0]above:$p_1$:}] (p1) {
            \node {$a_{21}$}; & \node {$a_{24}$}; \\
            \node {$a_{41}$}; & \node {$a_{44}$}; \\
        };
        \path (p0.east) ++(.25, 0) node[matrix, plot3, anchor=west, label={[inner sep=0]above:$p_2$:}] (p2) {
            \node {$a_{12}$}; & \node {$a_{15}$}; \\
            \node {$a_{32}$}; & \node {$a_{35}$}; \\
        };
        \path (p1.east) ++(.25, 0) node[matrix, plot4, anchor=west, label={[inner sep=0]above:$p_3$:}] (p3) {
            \node {$a_{22}$}; & \node {$a_{25}$}; \\
            \node {$a_{42}$}; & \node {$a_{45}$}; \\
        };
        \path (p2.east) ++(.25, 0) node[matrix, plot5, anchor=west, label={[inner sep=0]above:$p_4$:}] (p4) {
            \node {$a_{13}$}; \\
            \node {$a_{33}$}; \\
        };
        \path (p3.east) ++(.25, 0) node[matrix, plot6, anchor=west, label={[inner sep=0]above:$p_5$:}] (p5) {
            \node {$a_{23}$}; \\
            \node {$a_{43}$}; \\
        };

        \path (p0.west) -- (p1.west) coordinate[pos=.5] (m);
        \path (m) ++(-.75, 0) node[matrix, anchor=east] (dist) {
            \node[plot1] {$a_{11}$}; & \node[plot3] {$a_{12}$}; & \node[plot5] {$a_{13}$}; & \node[plot1] {$a_{14}$}; & \node[plot3] {$a_{15}$}; \\
            \node[plot2] {$a_{21}$}; & \node[plot4] {$a_{22}$}; & \node[plot6] {$a_{23}$}; & \node[plot2] {$a_{24}$}; & \node[plot4] {$a_{25}$}; \\
            \node[plot1] {$a_{31}$}; & \node[plot3] {$a_{32}$}; & \node[plot5] {$a_{33}$}; & \node[plot1] {$a_{34}$}; & \node[plot3] {$a_{35}$}; \\
            \node[plot2] {$a_{41}$}; & \node[plot4] {$a_{42}$}; & \node[plot6] {$a_{43}$}; & \node[plot2] {$a_{44}$}; & \node[plot4] {$a_{45}$}; \\
        }; 

        \path (p0.north) -- (p4.north) coordinate[pos=.5] (m);
        \path (m) ++(0, .5) node (ld) {local data:};
        \node at (dist |- ld) {distributed matrix:};

        \makeatletter
        \newcommand{\vast}{\bBigg@{4}}
        \makeatother

        \node[inner sep=0] at (dist.west) {$\vast($};
        \node[inner sep=0] at (dist.east) {$\vast)$};

        \foreach \i in {0,...,5} {
            \node[inner sep=0] at (p\i.west) {$\bigg($};
            \node[inner sep=0] at (p\i.east) {$\bigg)$};
        }
    \end{tikzpicture}
    
    \caption{
        Default 2D matrix distribution on a $2 \times 3$ process grid.
    }
    \label{fig:elem2D}
\end{figure}
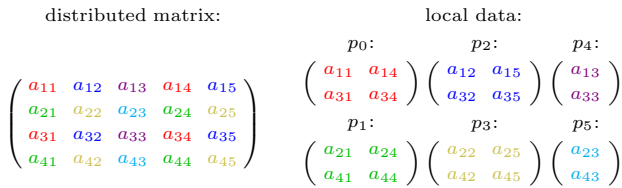

As a framework for distributed-memory dense linear algebra operations, we use 
{\sc Elemental}~\cite{elemental}. This is a C++ library, based on the 
Message Passing Interface ({\sc MPI}), 
that operates on a virtual two-dimensional grid of processes; the name is
inspired by the fact that, in general,
matrices are cyclically distributed across a 2D grid of processes in an
element-wise fashion.
This principal distribution\footnote{In {\sc Elemental}'s notation: $[MC, MR]$.} is shown in~\autoref{fig:elem2D}; 

Algebraic operations on distributed matrices involve two stages:
data redistribution (communication), and invocation of single-node
{\sc BLAS} or {\sc LAPACK} routines (computation). 
Optimal performance is 
attained by minimizing communication within the redistributions.  
In most cases, as shown in~\cite{elemental}, 
this is achieved by choosing the process grid to be as close to
a perfect square as possible.

While a square process grid is optimal for performance, since all processes
only hold non-contiguous portions of the matrix, it complicates loading
contiguously stored data from files into a distributed matrix.  In the context
of GWAS, the algorithm has to load two objects of different nature: the matrix $M$, and
the collections of vectors $X_{blk}$; the special nature of the latter determines
that the vectors can be loaded and processed in any order.

For loading $M$, we first read contiguous panels into the local memory
of each process via standard file operations,
and then construct the global (distributed) version of $M$ by accumulating 
the panels. This is done via {\sc Elemental}'s
axpy-interface, a feature that makes it possible to add 
node-local matrices to a global one. 

\begin{figure}[t]
    \scriptsize
    \centering
    \tikzset{external/export=true}

    \begin{tikzpicture}
        \node[matrix, plot1, label={[inner sep=0]above:$p_0$:}] (p0) {
            \node {$a_{11}$}; & \node {$a_{17}$}; \\
            \node {$a_{21}$}; & \node {$a_{27}$}; \\
            \node {$a_{31}$}; & \node {$a_{37}$}; \\
            \node {$a_{41}$}; & \node {$a_{47}$}; \\
        };
        \path (p0.east) ++(.25, 0) node[matrix, plot3, anchor=west, label={[inner sep=0]above:$p_2$:}] (p1) {
            \node {$a_{12}$}; & \node {$a_{18}$}; \\
            \node {$a_{22}$}; & \node {$a_{28}$}; \\
            \node {$a_{32}$}; & \node {$a_{38}$}; \\
            \node {$a_{42}$}; & \node {$a_{48}$}; \\
        };
        \path (p1.east) ++(.25, 0) node[matrix, plot5, anchor=west, label={[inner sep=0]above:$p_4$:}] (p2) {
            \node {$a_{13}$}; \\
            \node {$a_{23}$}; \\
            \node {$a_{33}$}; \\
            \node {$a_{43}$}; \\
        };
        \path (p2.east) ++(.25, 0) node[matrix, plot2, anchor=west, label={[inner sep=0]above:$p_1$:}] (p3) {
            \node {$a_{14}$}; \\
            \node {$a_{24}$}; \\
            \node {$a_{34}$}; \\
            \node {$a_{44}$}; \\
        };
        \path (p3.east) ++(.25, 0) node[matrix, plot4, anchor=west, label={[inner sep=0]above:$p_3$:}] (p4) {
            \node {$a_{15}$}; \\
            \node {$a_{25}$}; \\
            \node {$a_{35}$}; \\
            \node {$a_{45}$}; \\
        };
        \path (p4.east) ++(.25, 0) node[matrix, plot6, anchor=west, label={[inner sep=0]above:$p_5$:}] (p5) {
            \node {$a_{16}$}; \\
            \node {$a_{26}$}; \\
            \node {$a_{36}$}; \\
            \node {$a_{46}$}; \\
        };

        \path (p0.north) -- (p5.north) coordinate[midway] (m);
        \path (m) ++(0, .66) node[matrix, anchor=south, label={above:distributed matrix:}] (dist) {
            \node[plot1] {$a_{11}$}; & \node[plot3] {$a_{12}$}; & \node[plot5] {$a_{13}$}; & \node[plot2] {$a_{14}$}; & \node[plot4] {$a_{15}$}; & \node[plot6] {$a_{16}$}; & \node[plot1] {$a_{17}$}; & \node[plot3] {$a_{18}$}; \\
            \node[plot1] {$a_{21}$}; & \node[plot3] {$a_{22}$}; & \node[plot5] {$a_{23}$}; & \node[plot2] {$a_{24}$}; & \node[plot4] {$a_{25}$}; & \node[plot6] {$a_{26}$}; & \node[plot1] {$a_{27}$}; & \node[plot3] {$a_{28}$}; \\
            \node[plot1] {$a_{31}$}; & \node[plot3] {$a_{32}$}; & \node[plot5] {$a_{33}$}; & \node[plot2] {$a_{34}$}; & \node[plot4] {$a_{35}$}; & \node[plot6] {$a_{36}$}; & \node[plot1] {$a_{37}$}; & \node[plot3] {$a_{38}$}; \\
            \node[plot1] {$a_{41}$}; & \node[plot3] {$a_{42}$}; & \node[plot5] {$a_{43}$}; & \node[plot2] {$a_{44}$}; & \node[plot4] {$a_{45}$}; & \node[plot6] {$a_{46}$}; & \node[plot1] {$a_{47}$}; & \node[plot3] {$a_{48}$}; \\
        }; 

        \path (m) ++(0, .5) node (ld) {local data:};

        \makeatletter
        \newcommand{\vast}{\bBigg@{4}}
        \makeatother

        \node[inner sep=0] at (dist.west) {$\vast($};
        \node[inner sep=0] at (dist.east) {$\vast)$};

        \foreach \i in {0,...,5} {
            \node[inner sep=0] at (p\i.west) {$\vast($};
            \node[inner sep=0] at (p\i.east) {$\vast)$};
        }
    \end{tikzpicture}
    
    \caption{
        1D matrix distribution on a $1 \times 6$ process grid.
    }
    \label{fig:elem1D}
\end{figure}
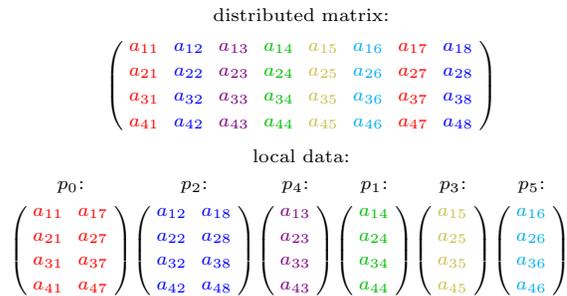

For loading $X_{blk}$ instead, 
a collection of contiguously stored vectors
is read into memory through more efficient means than the axpy-interface by
exploiting that, as long as consistently handled, the order of the vectors is
irrelevant.  The trick is to use a matrix that is distributed on a virtual 1D
reordering of the grid into a row of processes.  As shown in
\autoref{fig:elem1D}, the process-local data of such a matrix is a set of full
columns, which can be loaded from a contiguous data-file.  While these local
columns are not adjacent in the distributed matrix,
{\sc Elemental} guarantees that
all algebraic operations
performed on them maintain their order. 
For performance reasons, prior to any computation,
the matrix on the 1D ordering of this grid needs to be redistributed
to conform to the initial 2D process grid (\autoref{fig:elem2D}).
This
redistribution, provided by {\sc Elemental}, can internally be
performed most efficiently through a single {\tt MPI\_Alltoall} if the 1D grid
is the concatenation of the rows of the 2D grid.\footnote{In {\sc Elemental}:
$[\ast, VR]$.}

\subsection{The parallel algorithm}
\begin{lstlisting}[
    float=t,
    label=alg:dist,
    caption={
        Distributed memory version of \autoref{alg:base}. 
        Asynchronous I/O operations are depicted
        \green{green}, 
        distributed matrices and operations in \blue{blue}, 
        and quantities that differ across processes in \red{red}.
        %\pdj{ $M$ has to be loaded too }
        %\ep{same for $X_L$ and $y$. Is this relevant for the algorithm?}
    }
]
load_start !first! $\red{X_{blk}}$
$\blue{L L^T \coloneqq M}$
$\blue{\overline X_L \coloneqq L^{-1} X_L}$, !%
!           $\blue{\overline y \coloneqq L^{-1} y}$
copy $\overline X_L\ \blue{\coloneqq \overline X_L}$, $\overline y\ \blue{\coloneqq \overline y}$
$S_{TL} \coloneqq \overline X_L^T \overline X_L$, !%
!           $\overline b_T \coloneqq \overline X_L^T y$
for each $blk$
    load_wait !current! $\red{X_{blk}}$
    if not !last! $blk$: load_start !next! $\red{X_{blk}}$
    set $\blue{X_{blk}} \coloneqq$ combine($\red{X_{blk}}$)
    $\blue{\overline X_{blk} \coloneqq L^{-1} X_{blk}}$
    set $\red{\overline X_{blk}} \coloneqq$ localpart($\blue{\overline X_{blk}}$)
    $\red{S_{blk}} \coloneqq \red{\overline X_{blk}}^T \overline X_L$
    for $i$ in  $\{1, \ldots, \frac{m_{blk}}{np}\}$
        set $\red{\overline X_{Ri}} \coloneqq \red{\overline X_{blk}}[i]$, $\red{S_{BLi}} \coloneqq \red{S_{blk}}[i]$
        $\red{S_{BRi}} \coloneqq \red{\overline X_{Ri}}^T \red{\overline X_{Ri}}$
        $\red{\overline b_{Bi}} \coloneqq \red{\overline X_{Ri}}^T \overline y$
        set $\red{S_i} \coloneqq \splittt{S_{TL}}{\ast}{\red{S_{BLi}}}{\red{S_{BRi}}}$, $\red{\overline b_i} \coloneqq \splitto{\overline b_T}{\red{\overline b_{Bi}}}$
        $\red{b_i} \coloneqq \red{S_i^{-1}} \red{\overline b_i}$
        set $\red{b_{blk}}[i] \coloneqq \red{b_i}$
    end
    if not !first! $blk$: store_wait !previous! $\red{b_{blk}}$
    store_start !current! $\red{b_{blk}}$
end
store_wait !last! $\red{b_{blk}}$
\end{lstlisting}

%%% Local Variables: 
%%% mode: latex
%%% TeX-master: "Main"
%%% End: 

In~\autoref{alg:dist}, we present the distributed-memory version 
of~\autoref{alg:ooc} for $np$ processes; 
the matrices that are distributed among the processes 
and the corresponding operations are highlighted in~\blue{blue};
the quantities that differ from one process to another are instead in \red{red}.

The algorithm begins (line {\tt 1}) by loading the first
$\frac{m_{blk}}{np}$ vectors $X_{Ri}$ into a local block
$\red{X_{blk}}$ on each process asynchronously.
It commences with the initially distributed
$\blue{M}$, $\blue{X_L}$, and $\blue{y}$, and computes
$\blue{L}$, $\blue{\overline X_L}$, and $\blue{\overline y}$
(lines {\tt 2} -- {\tt 3}). 
Then, $X_L$ and $y$, local copies of $\blue{X_L}$ and
$\blue{y}$, respectively, are created on each process (line {\tt 4}).
Since small local computations are significantly more efficient than the 
distributed counterparts, 
$S_{TL}$ and $b_T$ are computed redundantly
by all processes (line {\tt 5}).

In
line {\tt 9}, the asynchronously loaded blocks $\red{X_{blk}}$ are ---without any communication or memory 
transfers--- seen as the columns of $\blue{X_{blk}}$ that are cyclically distributed
across a 1D process grid as described in~\autoref{sec:elemental}. 
Since in {\sc Elemental}
matrix operations require all operands to be 
in the default distribution across the
2D~grid, $\blue{X_{blk}}$ and $\blue{\overline X_{blk}}$ are redistributed
before and after the computation in line {\tt 10}, respectively.  Once
$\blue{\overline X_{blk}}$ is computed and redistributed, in line {\tt 11},
each process views its local columns of this matrix as $\red{\overline
X_{blk}}$; since the distributions of $\blue{X_{blk}}$ and $\blue{\overline
X_{blk}}$ are identical, these ---without communication or 
transfers--- correspond to the columns of $\red{X_{blk}}$.

In addition to blocking $\red{X_{Ri}}$ and $\red{b_{Bi}}$, 
the computation of all row vectors $\red{S_{BLi}}$ belonging to the current
block is combined into a single matrix product (line {\tt 12}) resulting in the
$\red{S_{BLi}}$ being stacked in a block $\red{S_{blk}}$.
In line {\tt 14}, $\red{S_{BLi}}$ is selected from
$\red{S_{blk}}$, along with $\red{X_{Ri}}$ from $\red{X_{blk}}$ for the
innermost loop.  This loop then computes the local $\red{b_{blk}}$
independently on each process.  Finally, $\red{b_{blk}}$ (whose columns
$\red{b_i}$ corresponds to the initially loaded vectors $\red{X_{Ri}}$ within
$\red{X_{blk}}$) is stored asynchronously, while the next iteration commences.

%%% Local Variables: 
%%% mode: latex
%%% TeX-master: "Main"
%%% End: 

\subsection{Performance Results}
We compile \distooc, the C++-implementation of~\autoref{alg:dist},
with the GNU C compiler
(version~4.7.2), use {\sc Elemental} (version~0.78-dev) with {\sc OpenMPI}
(version~1.6.4) and link to Intel's Math Kernel Library (MKL version~11.0).
In our tests, we use a compute cluster with 40 nodes, each equipped 
with 16~GB of RAM and two
quad-core Intel Harpertown E5450 processors running at 3.00~Ghz.
The nodes are connected via InfiniBand and access
a high speed Lustre file system.

Throughout all our experiments, we use the empirically optimal local block-size
$\frac{m_{blk}}{np} = 256$ by choosing $m_{blk} = 256 np$.

\subsubsection{Processing huge numbers of SNPs out-of-core}
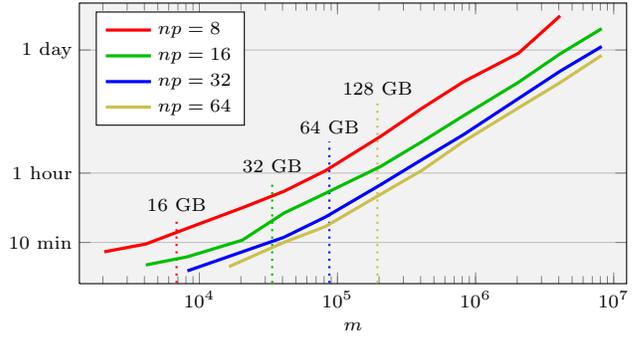
\begin{figure}[t]
    \scriptsize
    \centering
    \tikzset{external/export=true}

    \begin{tikzpicture}
        \begin{loglogaxis}[
            xlabel={$m$},
            ytick={60,600,3600,86400,604800},
            yticklabels={1 min, 10 min, 1 hour, 1 day, 1 week},
            legend pos=north west,
            every axis plot/.append style={
                very thick
            },
        ]
            \draw[plot1, thick, dotted] (axis cs:6841, 1) -- ++(0, 7) node[anchor=south, black] {16 GB};
            \draw[plot2, thick, dotted] (axis cs:33684, 1) -- ++(0, 8) node[anchor=south, black] {32 GB};
            \draw[plot3, thick, dotted] (axis cs:87368, 1) -- ++(0, 9) node[anchor=south, black] {64 GB};
            \draw[plot4, thick, dotted] (axis cs:194737, 1) -- ++(0, 10) node[anchor=south, black] {128 GB};
            \addplot[plot1] file {figures/data/m/8.dat};
            \addlegendentry{$np = 8$}
            \addplot[plot2] file {figures/data/m/16.dat};
            \addlegendentry{$np = 16$}
            \addplot[plot3] file {figures/data/m/32.dat};
            \addlegendentry{$np = 32$}
            \addplot[plot4] file {figures/data/m/64.dat};
            \addlegendentry{$np = 64$}
        \end{loglogaxis}
    \end{tikzpicture}

    \caption[Performance of \distooc ($m$)]{
      Performance of \distooc as a function of $m$.
      Here, $n = 40{,}000$, $p = 4$, and $m$ ranges from $2{,}048$ to $8.2 \cdot 10^6$.
      The vertical lines are
      limits for a theoretical in-core version of the parallel algorithm
      imposed by the accumulated RAM sizes.
    }
    
    \label{fig:dist:m}
    \tikzset{external/export=false}
\end{figure}

Since \distooc incorporates the double-buffering method introduced in
\autoref{sec:diego}, it can process datasets with arbitrarily large $m$
without introducing any overhead due to I/O operations.
To confirm this claim, we perform a series of experiments, using
 $np = 8$, $16$, $32$, and $64$
cores (1, 2, 4, and 8 nodes) to solve a system of size $n = 40{,}000$ and $p = 4$ with
increasing dataset size $m$.
The performance of these experiments is presented in \autoref{fig:dist:m},
where the vertical lines mark the points at which the 16~GB of RAM per
node are insufficient to store all $m$ vectors $X_{Ri}$.
The plot shows a very smooth behavior with $m$ (dominated by the
triangular solve in \autoref{alg:dist}, line {\tt 10}) well beyond this in-core memory limit.

\subsubsection{Increasing the population size $n$}
\begin{figure}[t]
    \scriptsize
    \centering
    \tikzset{external/export=true}

    \begin{tikzpicture}
        \begin{axis}[
            xlabel={$n$},
            ylabel={time [hours]},
            ymax=4,
            xmax=1e5,
            legend pos=north west,
            every axis plot/.append style={
                very thick
            },
            xtick={0,20000,40000,60000,80000,100000},
            xticklabels={$0$,$20{,}000$,$40{,}000$,$60{,}000$,$80{,}000$,$10^5$},
        ]
            \draw[plot1, very thick, dotted] (axis cs:43369, 0) -- (axis cs:43369, 1.2) node[anchor=south, black] {16 GB};
            \draw[plot2, very thick, dotted] (axis cs:59677, 0) -- (axis cs:59677, 2)   node[anchor=south, black] {32 GB};
            \draw[plot3, very thick, dotted] (axis cs:81203, 0) -- (axis cs:81203, 3.2) node[anchor=south, black] {64 GB};

            \addplot[plot1] file {figures/data/n/8.dat};
            \addlegendentry{$np = 8$}
            \addplot[plot2] file {figures/data/n/16.dat};
            \addlegendentry{$np = 16$}
            \addplot[plot3] file {figures/data/n/32.dat};
            \addlegendentry{$np = 32$}
            \addplot[plot4] file {figures/data/n/64.dat};
            \addlegendentry{$np = 64$}
        \end{axis}
    \end{tikzpicture}

    \caption[Performance of \distooc ($n$)]{
        Performance of \distooc as a function of $n$.  
        $p = 4$, $m = 65{,}536$, and $n$ ranges from $5{,}000$ to $100{,}000$.
        The vertical lines indicate the
        limits imposed by the accumulated RAM sizes.
    }
    \label{fig:dist:n}
    \tikzset{external/export=false}
\end{figure}
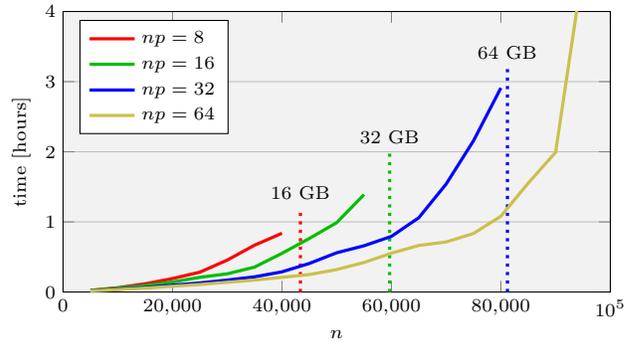

We now turn
to the main goal of our effort: performing computations on systems
whose matrix $M \in \mathbb R^{n \times n}$ exceeds the capacity of the main
memory.  For this purpose, we use $m = 65{,}536$, $p = 4$ and execute \distooc
on $np = 8$, $16$, $32$, and $64$ cores (1,
2, 4, and 8 nodes) with increasing matrix size $n$. \autoref{fig:dist:n}
reports the performance of these executions, which is dominated by the cubic
complexity of the Cholesky factorization of $\blue{M}$ (\autoref{alg:dist}, line
{\tt 2}).
The vertical lines indicate where
the nodes' memory would be exceeded by the size of the distributed $\blue{M}$
and the buffers for $\blue{X_{blk}}$.
The plot shows that our implementation succeeds in overcoming these memory
limitations through increasing the number of nodes.  

\subsubsection{Strong scalability}
\label{sec:strong}
\begin{figure}[t]
    \scriptsize
    \centering
    \tikzset{external/export=true}

    \begin{tikzpicture}
        \begin{axis}[
            xlabel={$np$},
            ylabel={time [min]},
            legend pos=north east,
            xmin={},
            xtick={8,16,24,32,48,64},
            every axis plot/.append style={
                very thick
            },
        ]
            \addplot[plot1] file {figures/data/strong/30000.49152.dat};
            \addlegendentry{$n = 30{,}000$, $m = 49{,}152$}
            \addplot[plot2] file {figures/data/strong/30000.98304.dat};
            \addlegendentry{$n = 30{,}000$, $m = 98{,}304$}
            \addplot[plot3] file {figures/data/strong/40000.49152.dat};
            \addlegendentry{$n = 40{,}000$, $m = 49{,}152$}
            \addplot[plot4] file {figures/data/strong/40000.98304.dat};
            \addlegendentry{$n = 40{,}000$, $m = 98{,}304$}
            \addlegendimage{gray, dotted}
            \addlegendentry{optimal scaling}
            \addplot[plot1, dotted, domain=8:64] {28.1901666667 / (x / 8)};
            \addplot[plot2, dotted, domain=8:64] {46.0605 / (x / 8)};
            \addplot[plot3, dotted, domain=8:64] {45.5353333333 / (x / 8)};
            \addplot[plot4, dotted, domain=8:64] {92.1773333333 / (x / 8)};
        \end{axis}
    \end{tikzpicture}

    \caption[Performance of \distooc ($np$)]{
        Performance of \distooc as a function of $np$.
        $p = 4$, and $np$ ranges from $8$ to $64$.
    }
    \label{fig:dist:strong}
    \tikzset{external/export=false}
\end{figure}
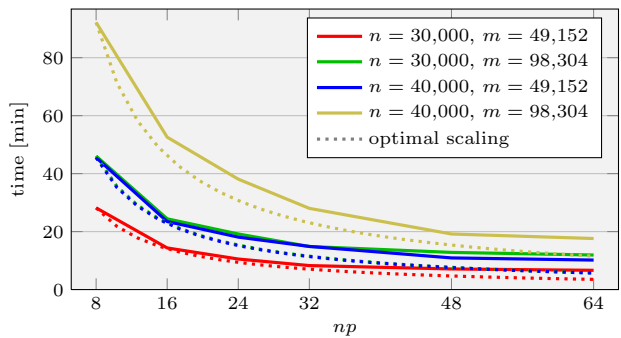

In practice, the problem sizes are bound to the specific GWAS and the interest
lies in solving \autoref{eq:GWAS} as fast as possible.
In the following experiment, we investigate how the time to solution is
reduced by \distooc through increasing the number of compute units, while
keeping the problem size constant.
In \autoref{fig:dist:strong}, we present the performance attained for four different problem sizes
with $8$ up to $64$ cores (1 through 8 nodes).
It shows perfect scalability for increasing the number of processes from $8$ to
$16$, reducing the runtime by a factor of 2.  With even more processes, the
parallel efficiency decreases, since the local portions of $\blue{L}$ become too small,
 but execution time is reduced further.

%%% Local Variables: 
%%% mode: latex
%%% TeX-master: "Main"
%%% End: 

\section{Conclusion}
We presented two parallel algorithms for solving the generalized least squares
problems that arise in genome-wide association studies (GWAS).  They address
the issue of growing dataset sizes due to the number of studied
polymorphisms $m$ and/or the population size $n$.

The first algorithm uses a double buffering
technique in order to process datasets with arbitrarily large numbers of
genetic polymorphisms.  Compared to other wide-spread GWAS-codes,
this algorithm's shared memory implementation, \smpooc,  was shown to be at least one order
of magnitude faster.

The second algorithm enables the processing of datasets involving large populations by
storing the covariance matrix in the combined main memory of distributed memory
architectures. \distooc, the implementation of this algorithm, was shown to scale
very well in both the population size and the number of processes used.

Together, these two algorithms form a viable basis for the challenges posed by
the scale of current and future genome-wise association studies.

\subsection{Future Work}
The work presented in this paper can be extended in several ways.
\begin{itemize}
    \item Hybrid parallelism, i.e., using multithreaded {\sc BLAS} and {\sc
    LAPACK}, as well as {\sc OpenMP}, offers further potential to boost the
    performance and efficiency of our distributed memory implementation
    \distooc.

    \item When a GWAS is interested in more than one trait $y$, a further
    dimension $j$ is added to the set of generalized least squares problems
    in \autoref{eq:GWAS}:
    $$ b_{ij} = \bigl(X_i^T M_j^{-1} X_i\bigl)^{-1} X_i^T M_j^{-1} y_j$$
    with $i = 1, \ldots, m$ and $j = 1, \ldots, t$.  A highly efficient shared
    memory implementation for this problem is already presented in~\cite{diego2d};
    only a distributed memory implementation on the lines of \distooc would be
    capable of solving this problem for large population sizes.

    \item Since the covariance matrix $M$ represents the relatedness of a
    diverse population, its few significant entries can be grouped close to the
    diagonal.  This allows to significantly reduce computation time by
    operating on banded matrices.
\end{itemize}

%%% Local Variables: 
%%% mode: latex
%%% TeX-master: "Main"
%%% End: 

\bibliographystyle{abbrv}
\bibliography{references}

\end{document}